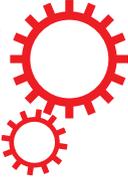

# Scalable Generation of Multi-mode NOON States for Quantum Multiple-phase Estimation

Lu Zhang & Kam Wai Clifford Chan

**Multi-mode NOON states have been attracting increasing attentions recently for their abilities of obtaining supersensitive and superresolved measurements for simultaneous multiple-phase estimation. In this paper, four different methods of generating multi-mode NOON states with a high photon number were proposed. The first method is a linear optical approach that makes use of the Fock state filtration to reduce lower-order Fock state terms from the coherent state inputs, which are jointly combined to produce a multi-mode NOON state with the triggering of multi-fold single-photon coincidence detections (SPCD) and appropriate postselection. The other three methods (two linear and one nonlinear) use *N*-photon Fock states as the inputs and require SPCD triggering only. All of the four methods can theoretically create a multi-mode NOON state with an arbitrary photon number. Comparisons among these four methods were made with respect to their feasibility and efficiency. The first method is experimentally most feasible since it takes considerably fewer photonic operations and, more importantly, requires neither the use of high-*N* Fock states nor high-degree of nonlinearity.**

NOON states are of great importance in quantum metrology, a field that studies the ultimate precision of the measurement of unknown physical parameters limited by the laws of quantum physics. It has been proven that NOON states can approach the Heisenberg limit (HL) for phase estimation, which gives a $\sqrt{N}$ gain in measurement sensitivity over the standard quantum limit (SQL) of using classical light sources only[1], where $N$ is the number of photons used. This is extremely meaningful for applications such as biological microscopy, where the imaging target is sensitive to illumination, that one can achieve the enhanced precision using as few photons as possible. Plenty of studies have been done in the generation of two-mode NOON states, both theoretically[2–12] and experimentally[13–15]. Their ability to achieve supersensitivity[16,17] and superresolution[14,18] in the single parameter estimation regime has also been experimentally demonstrated.

Recently, there is growing interest in the simultaneous estimation of multiple parameters using multi-mode quantum probing states[19–27]. The motivations can be summed up in two aspects: On one hand, multi-parameter estimation is not a trivial generalization of single-parameter estimation, in which the enhanced quantum limit is always attainable. For the multi-parameter scenario, the saturations of the optimal measurements of different parameters may not be attained simultaneously in general, since these measurements do not necessarily commute. Hence, multi-parameter estimation is an interesting topic for the study of quantum limit, which gives a metric on the multi-partite quantum states, and it can benefit many novel studies on quantum information theory, noncommutative geometry, etc. On the other hand, applications of quantum metrology, such as microscopy, quantum sensing, and quantum tomography, essentially involve multiple parameters that need to be estimated simultaneously. Specifically, for a phase imaging problem in a discretized multi-parameter model, multi-mode NOON states have been shown to have the ability of achieving super-resolution and super-sensitivity in simultaneous estimation with the potential advantage in estimation efficiency compared with the individual estimation counterpart[19]. Although the properties of the multi-mode NOON state were discussed in those previous studies, as far as the authors know, no known method of generating the NOON state with more than two modes and two photons has been presented.

In this paper, three linear and one nonlinear methods of generating *d*-mode *N*-photon NOON state are discussed in detail. The quantum state takes the form

School of Electrical and Computer Engineering, University of Oklahoma–Tulsa, Tulsa, Oklahoma, 74135, USA. Correspondence and requests for materials should be addressed to L.Z. (email: lu@ou.edu)

$$|\text{NOON}\rangle_d = \frac{1}{\sqrt{d}}(|N0\cdots0\rangle \pm |0N0\cdots0\rangle \pm \cdots \pm |0\cdots0N\rangle)_{1\cdots d}. \tag{1}$$

A distinguishing feature among these four methods is the quantum resource used as their inputs. They are respectively (i) coherent states and single photons, (ii) an evenly-distributed $N$-photon state and single photons, (iii) $d$ $N$-photon Fock states, and (iv) an $N$-photon Fock state and single photons. It is remarked that all four methods can theoretically produce NOON state with arbitrary photon number $N$, provided that the $N$-photon Fock states are available for the second, third and fourth methods. Method 4 additionally requires strong cross-Kerr nonlinearity. Deterministic generation of Fock states with six photons has been experimentally demonstrated using a superconducting quantum circuit[28]. Higher photon-number Fock states were shown theoretically to be achievable with methods such as the recycling strategies[29].

To compare the different NOON state generation methods, it will be useful to clarify the two general types of measurements commonly utilized in the studies of quantum information processing (QIP), namely preselection and postselection. QIP schemes based on preselection usually involve the explicit generation of the required quantum state, while those associated with postselection do not separate the required state from the undesired components until the final detection stage. Concretely for multiple phase estimation with the NOON state, by triggering on certain heralded modes, the preselection scheme extracts the NOON components from the photon sources and uses it to probe the target. That is, the NOON state is formed before interacting with the target. On the other hand, the postselection approach selects the useful NOON components carrying the phase information after the probing process, with both the NOON and non-NOON components present in the quantum state when it interacts with the target. Preselection may be the preferred method for various reasons (e.g., number of photons actually interacting with the target is exact) if it can be exploited efficiently, while in practice postselection is more commonly utilized in QIP experiments because of its relative ease of implementation using existing technology. Both methods are deemed effective when they can accomplish the same QIP task. In the discussion below, we regard the postselection method as effectively in generating the multi-mode NOON state.

## Results

In this section, the experimental setups of the four methods are described in detail. Particular attention is paid on calculating explicitly, for each method, the probability amplitude of the resulting NOON component compared with the input quantum state, which determines the intrinsic efficiency (generation probability) of the scheme.

**Method 1: Linear generation using coherent states and postselection.** The first method we proposed here uses coherent states and single photon states as the photon resources, where the latter can be generated almost perfectly using quantum dots[30]. Experimentally, a coherent state $|\alpha\rangle$ is a mathematical characterization of the output of an ideal single-mode laser, and it can be represented as a superposition of different Fock state components with the photon number following the Poisson distribution, i.e., $|\alpha\rangle = \exp(-|\alpha|^2/2)\sum_{n=0}^{\infty}\alpha^n|n\rangle/\sqrt{n!}$. The main idea of this method is to first generate a coherent mixture of $d$ identical coherent states $|\alpha, \ldots, \alpha\rangle_{1\cdots d}$, and then apply multiple Fock state filters (FSF)[31] on each $|\alpha\rangle$, where each FSF unit has the ability of filtering out any Fock state component $|k\rangle$ from the incident state by choosing the transmissivity of the beam splitter (BS) inside to be $T = k/(k+1)$. For each coherent state $|\alpha\rangle$, after the application of $\lfloor N/2 \rfloor$ FSFs with different BS transmissivity, the Fock components $|1\rangle, |2\rangle, \ldots, |\lfloor N/2 \rfloor\rangle$ can be cancelled out, where $\lfloor \cdot \rfloor$ is the floor function. Finally, triggered on the postselection of total $N$ photons terms in all $d$ modes in the final detection stage, the output state is essentially a $d$-mode NOON state, since all the $N$ photons could only come from any single mode of the $d$ coherent input states. Its experimental setup is sketched in Fig. 1(a), composed of $M = \lfloor N/2 \rfloor$ basic blocks of FSFs (green boxes).

Before mathematically showing its validity, we first briefly explain the working mechanism of the important unit–the Fock state filter–for future convenience, as illustrated in Fig. 1(b). It is adopted to filter out any certain Fock state component from the incident beam taking advantage of quantum multi-photon interference, with the help of a single photon catalyst, a beam splitter with certain transmissivity, and a single photon detector. The proof is shown below. Considering an input state in mode 1 as an arbitrary coherent superposition of the Fock state components

$$|\phi\rangle_1 = \sum_{n=0}^{\infty} C_n |n\rangle_1 = \sum_{n=0}^{\infty} C_n \frac{a_1^{\dagger n}}{\sqrt{n!}}|0\rangle_1, \tag{2}$$

the output state after combining $|\phi\rangle_1$ with a single photon state $|1\rangle_2$ in mode 2 on a BS with transmissivity $T = \cos^2\theta$ is

$$\begin{aligned} U_{12}(\theta)|\phi\rangle_1|1\rangle_2 &= \sum_{n=0}^{\infty} \frac{C_n}{\sqrt{n!}} U_{12}(\theta) a_1^{\dagger n} a_2^{\dagger}|0,0\rangle_{12} \\ &= \sum_{n=0}^{\infty} \frac{C_n}{\sqrt{n!}} (\cos\theta a_1^{\dagger} + i\sin\theta a_2^{\dagger})^n (\cos\theta a_2^{\dagger} + i\sin\theta a_1^{\dagger})|0,0\rangle_{12}, \end{aligned} \tag{3}$$

where $U_{12}(\theta) = \exp[i\theta(a_1^{\dagger}a_2 + a_1 a_2^{\dagger})]$ is the unitary operator of the BS with respect to modes 1 and 2 with field operators $a_1$ and $a_2$. When there is one and only one photon detected at the detector D in mode 2, the un-normalized state in output mode 1 is calculated in its density operator form as

$$|\phi\rangle'_1\langle\phi| \propto \mathrm{Tr}_2[|1\rangle_2\langle 1|(U_{12}(\theta)|\phi\rangle_1|1\rangle_2)\,(U_{12}(\theta)|\phi\rangle_1|1\rangle_2)^\dagger], \tag{4}$$

which is achieved by multiplying the single photon detection measurement $|1\rangle_2\langle 1|$ by the output state after the beam splitting, and then tracing off the measurement mode 2 since all measurement is destructive. This is equivalent to its state form

$$|\phi\rangle'_1 \propto \langle 1|_2 U_{12}(\theta)|\phi\rangle_1|1\rangle_2 = \sum_{n=0}^{\infty} C_n \cos^{n+1}\theta(1 - n\tan^2\theta)|n\rangle_1, \tag{5}$$

which is used in this paper to calculate the resulting NOON components explicitly. If the transmissivity of the BS is chosen to be $T = k/(k+1)$ (i.e., $\theta = \arctan 1/\sqrt{k}$), the probability of the $k$-photon component $|k\rangle_1$ appearing in the output state $|\phi\rangle'_1$ is zero. In other words, the $k$-photon Fock component is filtered out after the FSF, and the amplitudes of the other components $|n\rangle_1$ ($n \neq k$) are modulated by $\cos^{n+1}\theta(1 - n\tan^2\theta)$.

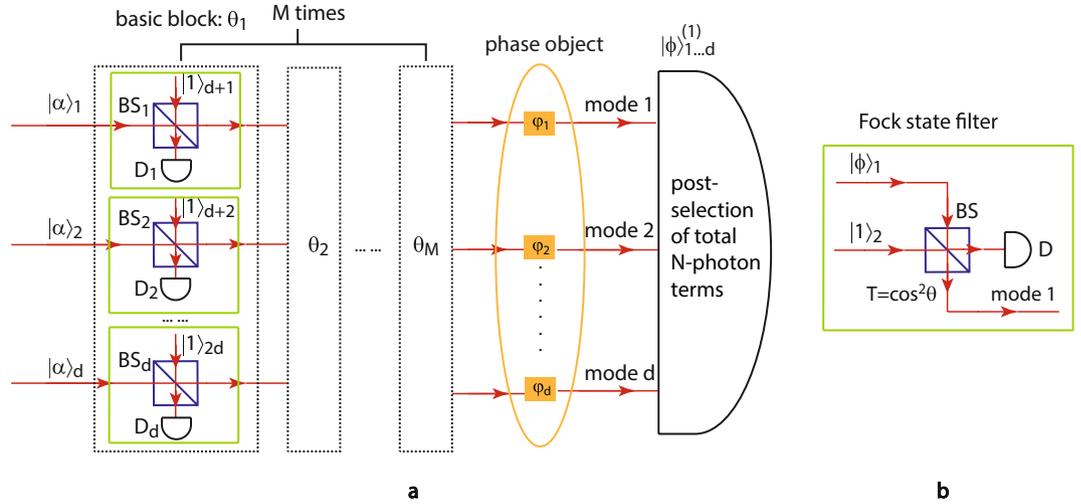

**Figure 1.** (**a**) Experimental setup of Method 1: generation of the $d$-mode $N$-photon NOON state using coherent states as input and postselection detection. In the scheme, $d$ identical coherent states jointly pass through $M$ basic blocks (shown as dashed boxes) sequentially. The $k$th basic block removes the $k$-photon terms in each mode of the inputs by Fock state filtration (green boxes) using $d$ identical beam splitters with transmissivity $\cos^2\theta_k = k/(k+1)$, $d$ single photons, and $d$ single photon detectors $\{D_j\}$ ($j = 1, \ldots, d$). For each mode, applying the basic block $M$ times with $\theta_k = \arctan(1/\sqrt{k})$ ($k = 1, 2, \ldots, M$) to the coherent state input results in a coherent-like state output with missing 1- to $M$-photon terms, when triggered on $M$-fold SPCD. A $d$-mode $N$-photon NOON state can then be obtained by choosing $M = \lfloor N/2 \rfloor$ and postselecting a total of $N$ photons in all of the output modes $1 \cdots d$ triggered on $dM$-fold SPCD after the quantum state interacts with the phase object. (**b**) Schematic of a Fock state filter. It is composed of a beam splitter with transmissivity $T = \cos^2\theta$, a single photon state $|1\rangle_2$ in one input mode, and a single photon detector D at one output. It takes an input state $|\phi\rangle_1$ with arbitrary photon statistics and filters out the $k$-photon component $|k\rangle_1$ from $|\phi\rangle_1$ by choosing BS transmissivity $T = k/(k+1)$ and triggering on a single photon detection at D.

More explicitly, the multi-mode NOON state is determined as follows. For the $k$th block, the state after adding $d$ single photons using $d$ beam splitters can be written as

$$\left[\bigotimes_{j=1}^{d} U^k_{j,d+j}\right]|\alpha, \ldots, \alpha\rangle_{1\cdots d}|1, \ldots, 1\rangle_{d+1,\ldots,2d}$$

$$= \bigotimes_{j=1}^{d} [U^k_{j,d+j}|\alpha\rangle_j|1\rangle_{d+j}]$$

$$= \bigotimes_{j=1}^{d} \left[ e^{-\frac{|\alpha|^2}{2}} \sum_{n_j=0}^{\infty} \frac{\alpha^{n_j}}{n_j!}(\cos\theta_k a_j^\dagger + i\sin\theta_k a_{d+j}^\dagger)^{n_j} \right.$$

$$\left. \times (\cos\theta_k a_{d+j}^\dagger + i\sin\theta_k a_j^\dagger)|0\rangle_{j,d+j} \right], \tag{6}$$

where $U^k_{j,d+j} = \exp[i\theta_k(a_j^\dagger a_{d+j} + a_j a_{d+j}^\dagger)]$ is the unitary operator of $BS_j$ with transmissivity $\cos^2\theta_k$. A $d$-fold SPCD at $\{D_j\}$ is then applied, projecting the state into

$$|\psi\rangle_{1\cdots d}^{(1),k} \propto \bigotimes_{j=1}^{d} \left\{ e^{-\frac{|\alpha|^2}{2}} \sum_{n_j=0}^{\infty} \frac{\alpha^{n_j}}{n_j!} [\cos^{n_j+1}\theta_k (1 - n_j \tan^2 \theta_k)] a_j^{\dagger n_j} |0\rangle_j \right\}. \tag{7}$$

Note that the right hand side of Eq. (7) is not normalized so as to show the probability amplitude relative to the input state explicitly, which enables the intrinsic efficiency of the method to be calculated later. Repeatedly applying this basic block $M$ times with different $\theta_k$ ($k = 1, \ldots, M$), the output state becomes

$$|\psi\rangle_{1\cdots d}^{(1)} \propto \bigotimes_{j=1}^{d} \left\{ e^{-\frac{|\alpha|^2}{2}} \sum_{n_j=0}^{\infty} \frac{\alpha^{n_j}}{n_j!} \left[ \prod_{k=1}^{M} \cos^{n_j+1}\theta_k (1 - n_j \tan^2 \theta_k) \right] a_j^{\dagger n_j} |0\rangle_j \right\}. \tag{8}$$

If $\theta_k$ is chosen to be $\theta_k = \arctan(1/\sqrt{k})$ (i.e., $\cos^2 \theta_k = k/(k+1)$), any term in $|\psi\rangle_{1\cdots d}^{(1)}$ with $n_j = k$ for any mode $j$ is canceled out. In other words, the 1- to $M$-photon terms in any of the $d$ modes disappear after the $M$ basic blocks, leaving the output state as

$$|\psi\rangle_{1\cdots d}^{(1)} \propto \bigotimes_{j=1}^{d} \left\{ e^{-\frac{|\alpha|^2}{2}} \left[ \frac{1}{\sqrt{M+1}} |0\rangle_j + \sum_{n_j=M+1}^{\infty} \frac{\alpha^{n_j}}{\sqrt{n_j!}} \left( \frac{1}{M+1} \right)^{\frac{n_j+1}{2}} \right.\right.$$
$$\left.\left. \times \frac{(n_j-1)!(-1)^M}{(n_j-M-1)!M!} |n_j\rangle_j \right] \right\}. \tag{9}$$

Finally, after the output state (9) is used to probe a target for multiple phase estimation, a postselection on exactly a total of $N$ photons in all the output modes $1\cdots d$ is performed as sketched in the right half of Fig. 1(a). Then only the NOON state components having all the $N$ photons in one mode $|N\rangle$ and no photons in any other mode $|0\rangle$ can contribute to the final detection. Therefore, we can write down the $d$-mode $N$-photon NOON state effectively generated upon the postselection and triggering as

$$|\phi\rangle_{1\cdots d}^{(1)} \propto c_1 (|N0\cdots 0\rangle + |0N0\cdots 0\rangle + \cdots + |0\cdots 0N\rangle)_{1\cdots d}, \tag{10}$$

where

$$c_1 = e^{-d\frac{|\alpha|^2}{2}} \frac{(-1)^M \alpha^N (N-1)!}{\sqrt{M+1}^{N+d} \sqrt{N!} (N-M-1)! M!}. \tag{11}$$

For the discussion on the efficiency of each method, we assume that all the detectors have unity efficiency and the beam splitters are lossless for simplicity. In this way, the intrinsic generation probability of the $d$-mode $N$-photon NOON state using the first method is

$$p_1 = d|c_1|^2 = \frac{d e^{-d|\alpha|^2} |\alpha|^{2N} (N-1)!}{(M+1)^{N+d} N (N-M-1)!^2 M!^2}, \tag{12}$$

which is a function of $|\alpha|^2$, $N$, and $d$. The generation probability $p_1$ is maximized at $|\alpha|^2_{\text{opt}} = N/d$ such that the mean total photon number of the input state is $N$:

$$p_1^{\text{opt}} = \frac{e^{-N} N^{N-2} N!}{d^{N-1} (M+1)^{N+d} (N-M-1)!^2 M!^2}. \tag{13}$$

As an example, for the 4-mode 4-photon NOON state generation with $M = 2$, the generation probability is approximately $4.2 \times 10^{-6}$.

**Method 2: Linear generation using an evenly-distributed N-photon input.** In this section, we presented a method of creating the $d$-mode $N$-photon NOON state using a $d$-mode evenly-distributed $N$-photon state and single photons, where the evenly-distributed $N$-photon state contains both NOON components with all $N$ photons being in one mode, and non-NOON components with $N$ photons distributed in multiple modes. Different from Method 1, this method does not require postselection, since its input has a deterministic number of photons. The main idea here is similar to that in Method 1, where the same Fock state filters are used to cancel out the non-NOON components from the system. Since the photon number of the input is fixed to be $N$, whenever the 1-photon term $|1\rangle$ is filtered out, the $(N-1)$-photon term $|N-1\rangle$ is discarded at the same time. Therefore, the non-NOON components with 1 to $(N-1)$ photons in any mode of the input are discarded as they pass through the same $M = \lfloor N/2 \rfloor$ basic blocks of FSFs, and only the NOON components survive eventually. This method was inspired from Zou et al.[3], where the two-mode NOON state was created in a similar way.

The required $d$-mode evenly-distributed $N$-photon state can be created by splitting an $N$-photon Fock state using $(d-1)$ beam splitters as shown in the green box in Fig. 2, where the $j$th BS ($U_j$) has transmissivity $T_j = 1/(d+1-j)$, ($j = 1, 2, \ldots, d-1$). Since a beam splitter introduces $\pi/2$ phase shift to the reflected beam, a phase shifter ($PS_j$) is applied to each mode to cancel out this effect. The unitary operation of the $d$ phase shifters is given by $U_{PS} = \prod_{j=1}^{d} \exp\left[-i\frac{\pi}{2} a_j^\dagger a_j (j-1)\right]$. Then the output after this process is an evenly-distributed $N$-photon state

$$\begin{aligned}|\psi\rangle^{(2)}_{1\cdots d} &= U_{PS} U_{d-1} \cdots U_1 |N\rangle_1 \\ &= \frac{1}{\sqrt{N!}\, d^{N/2}} (a_1^\dagger + a_2^\dagger + \cdots + a_d^\dagger)^N |0\rangle_{1\cdots d} \\ &= \frac{1}{d^{N/2}} \sum_{n_1+n_2+\cdots+n_d=N} \sqrt{C^N_{n_1,\ldots,n_d}} |n_1, \ldots, n_{d-1}, n_d\rangle_{1\cdots d} \\ &= |\psi\rangle^{NOON}_{1\cdots d} + |\psi\rangle^{non-NOON}_{1\cdots d},\end{aligned} \quad (14)$$

which contains both the NOON component (un-normalized)

$$|\psi\rangle^{NOON}_{1\cdots d} = \frac{1}{d^{N/2}} (|N0\cdots 0\rangle + \cdots + |0\cdots 0N\rangle)_{1\cdots d}, \quad (15)$$

and the non-NOON component (un-normalized)

$$|\psi\rangle^{non-NOON}_{1\cdots d} = \frac{1}{d^{N/2}} \sum_{\substack{n_1+n_2+\cdots+n_d=N,\\ n_1\neq N,\cdots,n_d\neq N}} \sqrt{C^N_{n_1,\ldots,n_d}} |n_1, \ldots, n_{d-1}, n_d\rangle_{1\cdots d}. \quad (16)$$

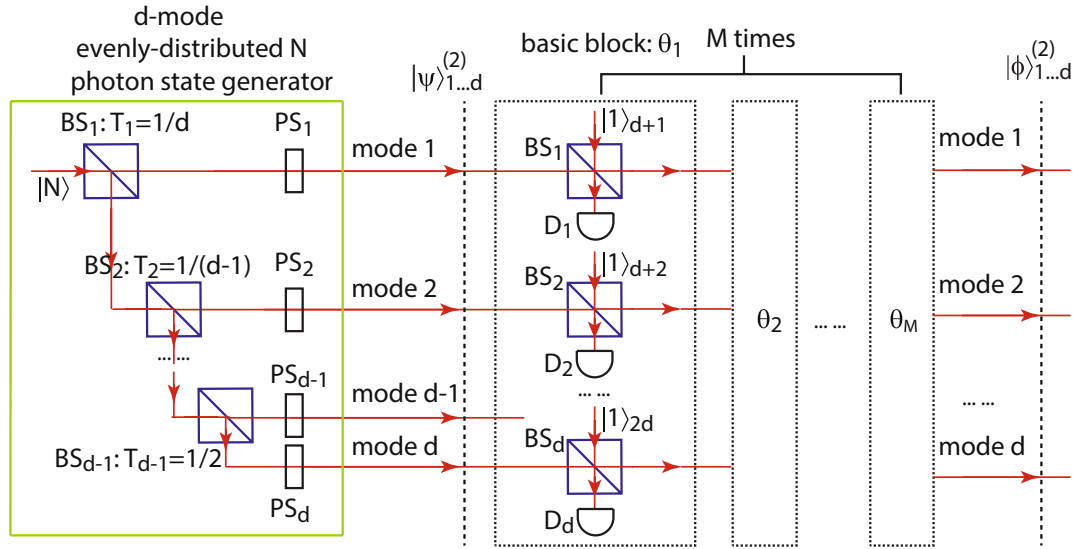

**Figure 2.** Experimental setup of Method 2: generation of the $d$-mode $N$-photon NOON state using a $d$-mode evenly-distributed $N$-photon state. The green box shows the creation of the required input, which is generated by splitting a Fock state $|N\rangle$ into $d$ beams using beam splitters $BS_j$ with transmissivity $T_j = 1/(d + 1 - j)$. The phase shifters $PS_j$ are chosen properly to cancel out the relative phase difference among the $d$ modes for the NOON state components. Then the state passes through $M = \lfloor N/2 \rfloor$ basic blocks (shown as dashed boxes) sequentially, each of which is identical to that in Method 1. A $d$-mode $N$-photon NOON state is resulted at the end of the $M$ basic blocks with the corresponding $\theta_k = \arctan(1/\sqrt{k})$ ($k = 1, 2, \ldots, M$).

The coefficient $C^N_{n_1,\ldots,n_d} = N!/(n_1! n_2! \ldots n_d!)$ denotes the multinomial distribution. This state is then fed into $M = \lfloor N/2 \rfloor$ basic blocks (the black dashed boxes), each of which is exactly the same as that in Method 1. The state after adding $d$ single photons in the $k$th block evolves into a $2d$-mode state

$$\begin{aligned}&\left[\bigotimes_{j=1}^{d} U^k_{j,d+j}\right] |\psi\rangle^{(2)}_{1\cdots d} |1,\cdots,1\rangle_{d+1,\ldots,2d} \\ &= \frac{1}{d^{N/2}} \sum_{n_1+n_2+\cdots+n_d=N} \frac{\sqrt{N!}}{n_1! n_2! \ldots n_d!} \left[\prod_{j=1}^{d} (\cos\theta_k a_j^\dagger + i\sin\theta_k a_{d+j}^\dagger)^{n_j} \times (\cos\theta_k a_{d+j}^\dagger + i\sin\theta_k a_j^\dagger)\right] |0\rangle_{1\cdots 2d}\end{aligned} \quad (17)$$

.

Then the $d$-fold SPCD at $\{D_j\}$ projects the state into

$$|\psi\rangle_{1\cdots d}^{(2),k} \propto \frac{1}{d^{N/2}} \sum_{n_1+n_2+\cdots+n_d=N} \sqrt{\frac{N!}{n_1!n_2!\cdots n_d!}} (\cos\theta_k)^{N+d}$$
$$\times \left[\prod_{j=1}^{d} (1 - n_j\tan^2\theta_k)\right] |n_1, \cdots, n_d\rangle_{1\cdots d}. \quad (18)$$

When this state $|\psi\rangle_{1\cdots d}^{(2)}$ passes through each block from $k=1$ to $k=M=\lfloor N/2 \rfloor$ with $\theta_k = \arctan(1/\sqrt{k})$, the $k$th block cancels out all the non-NOON components with $k$ or $(N-k)$ photons in any mode, since the total photon number in the system is fixed to be $N$. Eventually only the terms with $N$ photons in one mode and vacuum in all the other modes survive, which is essentially a $d$-mode $N$-photon NOON state:

$$|\phi\rangle_{1\cdots d}^{(2)} \propto c_2(|N0\cdots 0\rangle + |0N0\cdots 0\rangle + \cdots + |0\cdots 0N\rangle)_{1\cdots d}, \quad (19)$$

where

$$c_2 = \frac{1}{d^{N/2}} \prod_{k=1}^{M} [(\cos\theta_k)^{N+d}(1 - N\tan^2\theta_k)] = \frac{(-1)^M(N-1)!}{d^{N/2}\sqrt{M+1}^{N+d}(N-M-1)!M!}. \quad (20)$$

The intrinsic generation probability of the $d$-mode $N$-photon NOON state using this method is then

$$p_2 = d|c_2|^2 = \frac{(N-1)!^2}{d^{N-1}(M+1)^{N+d}(N-M-1)!^2 M!^2}. \quad (21)$$

When $d=N=4$, the generation probability using Method 2 is $2.1 \times 10^{-5}$, which is 5 times more efficient than method 1.

**Method 3: Linear generation using $d$ $N$-photon Fock states.** Based on the work of Kok et al.[4], a cascading method was proposed in this section, which uses $d$ $N$-photon Fock states to generate a $d$-mode $N$-photon NOON state with $d=2^n$ ($n = 1, 2, \ldots$). The reason it can only generate NOON states with certain mode number $d$ is that one needs to balance the amplitude for each NOON component, which also applies to Method 4 in the next section. The experimental setup consists of $(d-1)$ basic blocks, called the entanglement generators in this paper, aligned in a cascading configuration such that the output of the previous generator is injected into the next generator. Each generator creates a two-mode $N$-photon entangled state by repeatedly reducing one or two photons (depending on the parity of $N$) from either of the two $N$-photon Fock state inputs $|N, N\rangle$ without the knowledge of their originating modes, such that $N$ photons will be reduced either all from mode 1 or all from mode 2 after each generator. This method works differently for even-$N$ and odd-$N$, which is discussed in detail as follows.

We first describe the even-$N$ NOON state generation. The experimental setup is sketched in Fig. 3(a), where $(d-1)$ entanglement generators shown as black dashed boxes are arranged in a cascading setup (essentially a binary tree). Every entanglement generator is in turn composed of $N/2$ sub-blocks (the green boxes), each of which contains two identical beam splitters $BS_1$ and $BS_2$, a 50:50 beam splitter $BS_3$, two single photon detectors $D_1$ and $D_2$, and one phase shifter PS. The transmissivities of both $BS_1$ and $BS_2$ for the $k$th sub-block ($k=1, 2, \ldots, N/2$) are $T_k = (N-k)/(N-k+1)$, which are optimally chosen in order to split two photons off from the dual-Fock state input $|N, N\rangle_{ad}$ with the highest probability. Then the output modes $b'$ and $c'$ are recombined using the 50:50 beam splitter $BS_3$, whose output are measured by $D_1$ and $D_2$. Whenever a two-fold SPCD is measured at $D_1$ and $D_2$ (denoted as $_{b''c''}\langle 11|$), two photons are reduced either from mode $b'$ or mode $c'$ due to the two-photon quantum interference. Similarly, when $N/2$ sub-blocks with the corresponding $T_k$ and phase shift $\psi_k = 2\pi k/N$ at mode $c'$ are applied, the $N$ photons can be reduced either from mode $a$ or mode $d$, which can be explained using the mathematical formula $\prod_{k=1}^{N/2}(a^2 + e^{i2\psi_k}d^2) = a^N \pm d^N$ with abuse of notations. This state after the first entanglement generator can be expressed as:

$$|\phi\rangle_{12}^{(3,\text{even})} \propto \left[\bigotimes_{k=1}^{N/2} M_k\right]|N, N\rangle_{ad}, \quad (22)$$

where

$$M_k = {}_{b''c''}\langle 11|e^{i\psi_k c'^\dagger c'} U_k = \frac{i}{\sqrt{2}}({}_{b'c'}\langle 20| + e^{i2\psi_k}{}_{b'c'}\langle 02|) U_k \quad (23)$$

is the measurement operation performed by the $k$th sub-block, and $U_k = \exp[i\theta_k(a^\dagger b + ab^\dagger + c^\dagger d + cd^\dagger)]$ is the unitary operator for $BS_1$ and $BS_2$. Using mathematical induction, one can prove that the output state $|\phi\rangle_{12}^{(3,\text{even})}$ is

$$|\phi\rangle_{12}^{(3,\text{even})} \propto \left(\frac{-i}{2}\right)^{N/2}\left[\prod_{k=1}^{N/2} \sin^2\theta_k \cos^{2(N-k)}\theta_k(a^2 + e^{i2\psi_k}d^2)\right]|N, N\rangle_{ad}$$
$$= c_{3a}(|0N\rangle \pm |N0\rangle)_{12}, \quad (24)$$

where

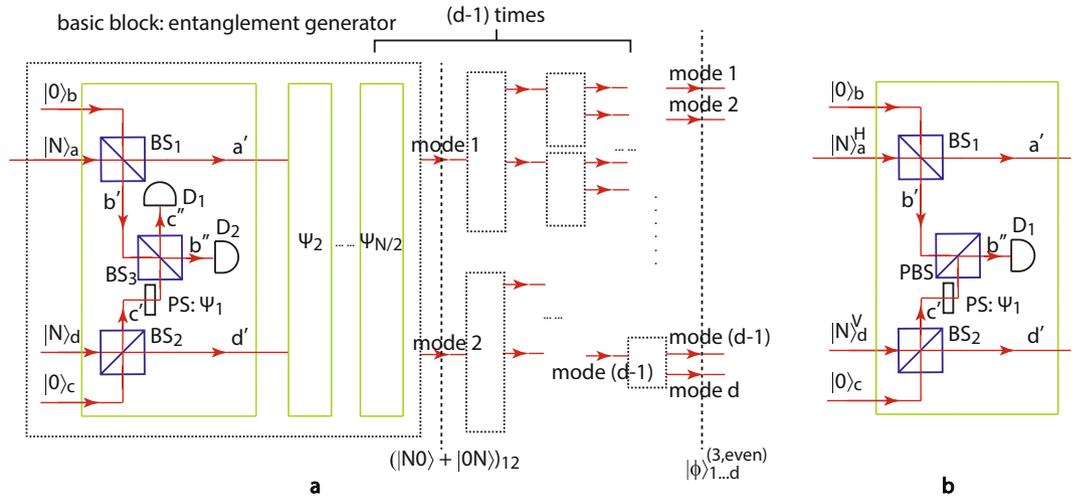

**Figure 3.** Experimental setup of Method 3: generation of the $d$-mode $N$-photon NOON state using $d$ Fock states as input. (**a**) Full setup with even $N$. A basic entanglement generator (black dashed box) creates two-mode entanglement when a Fock state $|N\rangle_a$ is fed into the system. The generator is composed of another Fock state input $|N\rangle_d$ and $N/2$ sub-blocks (green boxes), each of which reduces two photons from either of the two input modes without the knowledge of which mode they come from. This is achieved by splitting the two Fock states using two identical beam splitters $BS_1$ and $BS_2$ with transmissivity $T_k = (N-k)/(N-k+1)$ for the $k$th sub-block ($k = 1, 2, …, N/2$), then recombining mode $b'$ and $c'$ using a 50:50 beam splitter $BS_3$ with a phase shifter PS $\psi_k = 2\pi k/N$, and projecting out two photons using detectors $D_1$ and $D_2$. Applying the entanglement generator $(d-1)$ times with each input of the following generator aligning with one of the output modes of the previous one gives a $d$-mode $N$-photon NOON state. (**b**) The sub-block inside the entanglement generator for the case of odd $N$. To handle odd $N$, the photon polarization is exploited in the system, where two Fock states with horizontal and vertical polarizations are used. The transmissivities of $BS_1$ and $BS_2$ are chosen to be $T'_k = (2N - k)/(2N - k + 1)$ for the $k$th sub-block ($k = 1, 2, …, N$), $BS_3$ is replaced by a polarizing beam splitter PBS and a single photon detector $D_1$ is adopted in order to reduce one photon from either of the two input modes with a phase shifter $\psi_k = 2\pi k/N$.

$$c_{3a} = \frac{(-i)^{N/2}\sqrt{N!}}{2^N N^{N/2}}. \tag{25}$$

The upper (lower) sign in Eq. (24) applies to $N = 2 + 4q$ ($N = 4 + 4q$) with $q = 0, 1, 2, …$.

To extend this method to the multi-mode case, one can simply apply more entanglement generators whose input is aligned with one of the outputs of the previous block, as shown in Fig. 3(a). As an example, we align the output mode 2 of the first generator with the input of the next generator. In this case, together with another Fock state $|N\rangle_3$ from the second generator, the input then becomes a superposition of $|0NN\rangle_{123}$ and $|N0N\rangle_{123}$. Since the second generator only works on the state in modes 2 and 3, the state in mode 1 stays unchanged. Term $|0NN\rangle_{123}$ then creates another entanglement between modes 2 and 3 and evolves into $c_{3a}(|0N0\rangle_{123} + |00N\rangle_{123})$, while term $|N0N\rangle_{123}$ evolves into $c_{3b}|N00\rangle_{123}$ after $N$ photons are measured out from mode 3, with the coefficient

$$c_{3b} = \left(\frac{-i}{2}\right)^{N/2}\frac{\sqrt{N!}}{N^{N/2}}. \tag{26}$$

The three-mode state we obtain here is similar to a NOON state but with unbalanced amplitudes among different NOON components. In order to create balanced NOON states, the mode number has to be chosen as $d = 2^n$ ($n = 1, 2, …$). When this entanglement generating process is repeated $(d-1)$ times, the $d$-mode NOON state is finally obtained:

$$|\phi\rangle^{(3,\text{even})}_{1…d} \propto c_{3a}^{\log_2 d} c_{3b}^{d-\log_2 d - 1}(|N0···0\rangle \pm |0N0···0\rangle \pm ··· \pm |0···0N\rangle)_{1…d}, \tag{27}$$

whose generation probability is

$$p_3^{\text{even}} = d\left|c_{3a}^{\log_2 d} c_{3b}^{d-\log_2 d - 1}\right|^2 = \frac{1}{d^{N-1}}\left(\frac{N!}{2^N N^N}\right)^{d-1}. \tag{28}$$

This gives $p_3 = 3.1 \times 10^{-9}$ when $d = 4$ and $N = 4$.

For odd-$N$, we need to introduce a new degree of freedom, polarization, into the experiment in order to avoid the triggering of non-detection. The input state is now instead a dual-Fock states with orthogonal polarizations (e.g., horizontal and vertical). The main structure is still the same, except for some changes in the entanglement generator sub-block as shown in Fig. 3(b). Instead of reducing two photons, the *k*th sub-block ($k = 1, 2, \ldots, N$) only reduces one photon each time, where the transmissivities of $BS_1$ and $BS_2$ are chosen to be $T'_k = (2N-k)/(2N-k+1)$. $BS_3$ is replaced by a polarizing beam splitter PBS, which transmits horizontal photons and reflects vertical photons. A single photon detector is used to project one photon coming from either mode $b'$ or mode $c'$. Then with the same phase shift $\psi_k = 2\pi k/N$ ($k = 1, \ldots, N$), the output state after the first entanglement generator is a two-mode NOON state:

$$|\phi\rangle^{(3,\mathrm{odd})}_{12} \propto (-1)^N \left[\prod_{k=1}^{N} \sin\theta_k \cos^{(2N-k)}\theta_k (a + e^{i\psi_k}d)\right]|N, N\rangle_{ad}$$
$$= c'_{3a}(|0N\rangle \pm |N0\rangle)_{12}, \tag{29}$$

with $c'_{3a} = (-1)^N\sqrt{N!}/(2^N N^{N/2})$, where the upper (lower) sign applies to $N = 3 + 4q$ ($N = 5 + 4q$), with $q = 0, 1, 2, \ldots$. The remaining steps of the extension into multi-mode scenario are the same as those for even-*N*, and hence the odd-*N* scenario has the same intrinsic efficiency as in Eq. (28).

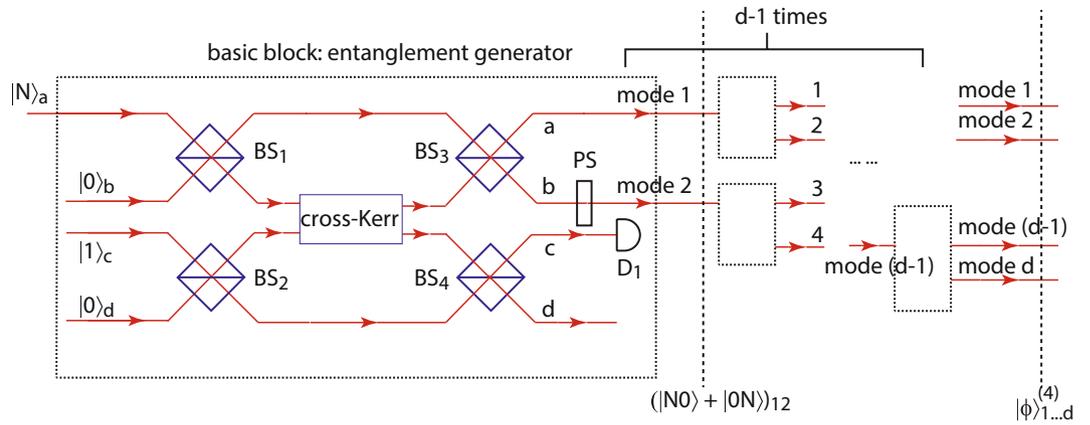

**Figure 4.** Experimental setup of Method 4: generation of the *d*-mode *N*-photon NOON state using cross-Kerr nonlinearity. The dashed box is the basic entanglement generator, which creates two-mode *N*-photon entanglement from a Fock state input and is made up of a cross-Kerr nonlinear medium, a single photon input, four 50:50 beam splitters $BS_j$ ($j = 1, 2, 3, 4$), a phase shifter PS with $\psi = -N\pi/2$, and a single photon detector $D_1$. Applying the basic block ($d-1$) times with each input of the following generator aligning with one of the output modes of the previous generator gives a *d*-mode *N*-photon NOON state.

**Method 4: Nonlinear generation using a cross-Kerr medium.** The last method presented here takes advantages of cross-Kerr nonlinearity suggested by Gerry *et al.*[5], where a cross-Kerr medium with nonlinearity degree $\chi = \pi$ is embedded into a Mach-Zehnder interferometer in order to act as an entanglement generator using an *N*-photon Fock state and a single photon state, and this process is shown in the dashed boxes in Fig. 4. In this paper, we extend this method to generate multi-mode NOON states with the mode number $d = 2^n$ ($n = 1, 2, \ldots$) by repeating this process ($d-1$) times in the same cascading configuration as that adopted in Method 3.

As shown in Fig. 4, the setup is composed of ($d-1$) entanglement generators (dashed boxes) applied in a cascading way, where the input of the next generator is aligned with one output of the previous generator. Mathematically, the cross-Kerr nonlinear effect in each generator can be represented by the unitary operator

$$U_K = e^{i\chi a^\dagger a d^\dagger d}, \tag{30}$$

where $\chi = \pi$ is the degree of nonlinearity required. The nonlinear medium is placed in an interferometer composed of a single photon state, four 50:50 beam splitters $BS_j$ ($j = 1, 2, 3, 4$), a phase shifter PS, and a single photon detector $D_1$. The parameters for the four beam splitters are chosen to be $\theta_1 = \theta_2 = -\theta_3 = -\theta_4 = \pi/4$. Then the output state after the four beam splitters and the cross-Kerr medium is given by

$$U^\dagger_{cd} U^\dagger_{ab} U_K U_{ab} U_{cd} |N, 0, 1, 0\rangle_{abcd}$$
$$= U^\dagger_{cd}\left(e^{\frac{i}{2}\chi d^\dagger d(a^\dagger a + b^\dagger b)} e^{\frac{1}{2}\chi d^\dagger d(ab^\dagger - a^\dagger b)}\right) U_{cd}|N, 0, 1, 0\rangle_{abcd}$$
$$= \frac{1}{2}[(|N0\rangle_{ab} + e^{i\frac{\pi}{2}N}|0N\rangle_{ab})|1, 0\rangle_{cd}$$
$$+ i(e^{i\frac{\pi}{2}N}|0N\rangle_{ab} - |N0\rangle_{ab})|0, 1\rangle_{cd}]. \tag{31}$$

| Method | 1 | 2 | 3 (even-$N$) | 4 |
|---|---|---|---|---|
| Major strength | coherent state input | preselection | preselection | high efficiency |
| Major weakness | postselection | Fock state input | low efficiency | strong nonlinearity |
| Generation Probability | $\frac{d^{-(N-1)}e^{-N}N^{N-2}N!}{(\lfloor\frac{N}{2}\rfloor+1)^{N+d}(\lceil\frac{N}{2}\rceil-1)!^2\lfloor\frac{N}{2}\rfloor!^2}$ | $\frac{d^{-(N-1)}(N-1)!^2}{(\lfloor\frac{N}{2}\rfloor+1)^{N+d}(\lceil\frac{N}{2}\rceil-1)!^2\lfloor\frac{N}{2}\rfloor!^2}$ | $\frac{d^{-(N-1)}N!^{d-1}}{(2^NN)^{d-1}}$ | $\frac{1}{d}$ |
| # of beam splitters | $d\lfloor N/2\rfloor$ | $d\lfloor N/2\rfloor + d - 1$ | $3N(d-1)/2$ | $4(d-1)$ |
| # of phase shifters | 0 | $d$ | $N(d-1)/2$ | $d-1$ |
| # of SPCD | $d\lfloor N/2\rfloor$ | $d\lfloor N/2\rfloor$ | $N(d-1)$ | $d-1$ |
| # of Fock states $|N\rangle$ | 0 | 1 | $d$ | 1 |
| # of single photons $|1\rangle$ | $d\lfloor N/2\rfloor$ | $d\lfloor N/2\rfloor$ | 0 | $d-1$ |

**Table 1.** Comparisons between the four multi-mode NOON state generation methods.

After applying a phase shift of $-N\pi/2$ to the output mode $b$, whenever a photon is detected at $D_1$, one can obtain a two-mode NOON state in modes 1 and 2:

$$|\phi\rangle^{(4)}_{12} \propto \frac{1}{2}(|N0\rangle_{12} + |0N\rangle_{12}). \tag{32}$$

Essentially, either the triggering of single photon detection in mode $c$ or mode $d$ can result in a two-mode NOON state. However, in order to extend this method to the multi-mode scenario, only one detector in mode $c$ is used in each entanglement generator as the preselection, since the single photon input can only trigger a detection in mode $c$ instead of mode $d$ when $|0\rangle_2$ component in $|\phi\rangle^{(4)}_{12}$ (Eq. (32)) is injected into the next generator. The scheme of adding more entanglement modes into the system is the same as that in Method 3. The final $d$-mode $N$-photon NOON state is then

$$|\phi\rangle^{(4)}_{1\cdots d} \propto \frac{1}{d}(|N0\cdots 0\rangle + |0N0\cdots 0\rangle + \cdots + |0\cdots 0N\rangle)_{1\cdots d}, \tag{33}$$

triggered on a $d$-fold SPCD. The generation probability for this method is

$$p_4 = \frac{1}{d}. \tag{34}$$

For the 4-mode 4-photon NOON state, $p_4 = 0.25$, which is much higher than the previous three methods. Nevertheless, the required high-degree of nonlinearity is extremely hard to achieve using current technologies.

**Discussion**
After describing the four methods of generating multi-mode NOON states in detail, we now make comparisons among them with respect to their feasibility and efficiency. In general, Methods 1 and 2 take advantage of the Fock state filtration, to reduce the unexpected components from the $d$-mode coherent states input and $d$-mode evenly-distributed $N$-photon state input, respectively. On the other hand, Methods 3 and 4 make use of multiple entanglement generators, each of which adds one mode entanglement into the system. The detail comparisons are shown in Table 1. Note that only the even-$N$ scenario for Method 3 is shown in the table, since the odd-$N$ and even-$N$ scenarios have the same efficiency and the only main difference between them is that the former requires twice the numbers of BSs and PSs.

Compared with Methods 2, 3 and 4 which require $N$-photon Fock state as inputs, Method 1 only uses coherent states as the input in addition to single photons. As high photon-number Fock states are relatively hard to obtain, Method 1 is thus most feasible among the four approaches. In addition, the number of optical components (such as BS and detectors) required in Method 1 are all in the degree of $O(dN/2)$, which is comparable with method 2, but much lower compared with Method 3. This is significant when the imperfect efficiency of the optical components is taken into account in practice. Due to the use of nondeterministic-photon-number input in Method 1, it requires postselection detection to discard all terms with photon number different from the intended one. Since current measurement methods are all destructive to photons, the postselection process would not be applied until the experiment using the NOON state finishes. On the contrary, the other three methods require preselection only.

Compared with Methods 1 and 2, both Methods 3 and 4 can only create the balanced NOON state with mode number $d = 2^n$ ($n = 1, 2, \ldots$), while the first two have no limitation on $d$. The last two methods use a similar cascading structure, making uses of $(d-1)$ entanglement generators, each of which creates a two-mode $N$-photon entangled state from separable input states. The difference between them lies in the entanglement creation method, where the former takes advantages of the 2-photon quantum interference, while the latter uses the nonlinear cross-Kerr medium. It appears that Method 4 requires much fewer optical components compared with the third one, however, the requirement of high nonlinearity degree still makes it the least feasible method to implement experimentally.

To compare the efficiency of the methods more explicitly, we plot the generation probabilities respectively for 4-photon NOON state with respect to mode number $d$ in Fig. 5(a) and 4-mode NOON state with respect to photon number $N$ in Fig. 5(b). In general, the generation probability decreases with increasing $d$ when $N$ is fixed, and vice versa. It is noticeable that the generation probability using Method 4 is much higher than that of the other three methods. It should however be noted that the comparison is made assuming the Fock states and strong cross-Kerr nonlinearity are available; the generation probability only reflects the proportion of the photons that are utilized in reference to all the photons that are used in the NOON state generation process. Method 2 has the higher generation probability among the three linear methods when $d \geq 3$, followed by Methods 1 and 3. Both Methods 1 and 2 are exponentially more efficient than Method 3 with either fixed-$d$ ($d \geq 3$) or fixed-$N$. The asymptotic ratio between the generation probabilities of Method 2 and Method 1 when $N$ approaches infinity is

$$\lim_{N \to \infty} \frac{p_2}{p_1^{\text{opt}}} = \lim_{N \to \infty} \frac{(N-1)! e^N}{N^{N-1}} = \sqrt{2\pi N}. \tag{35}$$

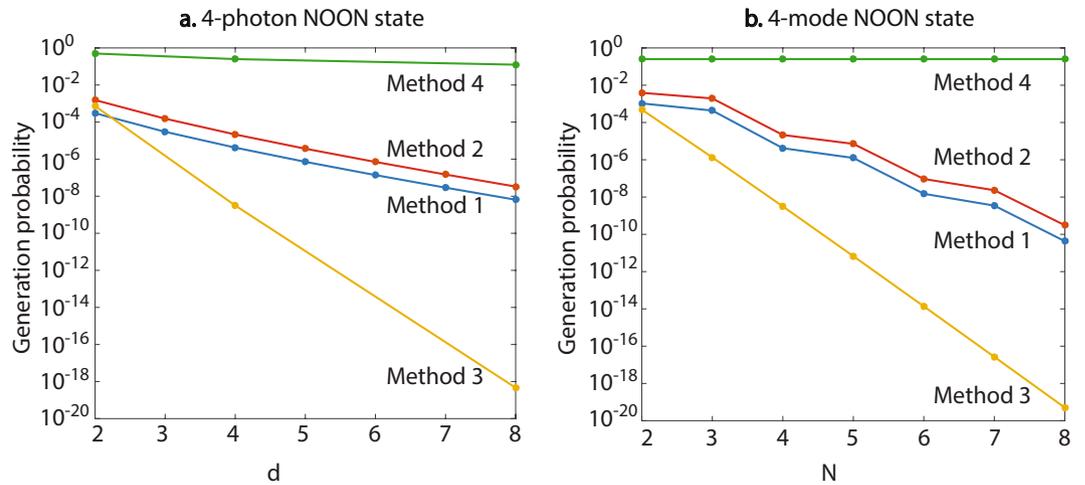

**Figure 5.** (**a**) Plots of the generation probabilities of 4-photon NOON state using the four methods with respect to the mode number $d$. Note that for the last two methods, only the points with $d = 2^n$ ($n = 1, 2, 3$) are plotted. (**b**) Plots of the generation probabilities of 4-mode NOON state with respect to the photon number $N$. The discrete points are connected for better visualization.

This means Method 2 is $\sqrt{2\pi N}$ times more efficient than Method 1 when $N$ is large.

The efficiency comparisons above are made under the assumptions that all the optical devices have no losses. However, in practice, losses always exist and need to be taken into consideration. For the first two methods, the main experimental challenge comes from the imperfection of the FSFs. The studies on FSF imperfection involve multiple variables, such as the input photon distributions, the single photon generation rate, the FSF sensitivity $k$ (used to filter out $|k\rangle$), and the photon detection sensitivity, where its efficiency cannot be simply characterized. In 2015, an experimental model of an imperfect FSF was proposed and studied from three aspects[32]: the use of photon detectors that cannot distinguish events between one photon detection and higher-photon-number detection, the imperfect single photon catalyst that is actually a superposition of a single photon state and a vacuum state, and mode mismatch between the input state and the single photon catalyst; and they calculated the success probability of an imperfect FSF that can filter out $|1\rangle$ from a coherent state using Eq. (14) in that paper[32]. This structure can be utilized to study the FSF efficiency $p_{\text{FSF}}(k)$ with other photon number sensitivity $k$, which is one of the next research directions based on the theoretical work here. The realistic efficiency of the third or fourth method also depends on the sensitivity of photon detectors $\eta_D$ and the generation efficiency of single photon states $\eta_{|1\rangle}$. Loosely speaking, the efficiency of each method considering the above two imperfections can be calculated by multiplying the corresponding theoretical efficiency $p$ by $\eta_D^x \eta_{|1\rangle}^y$, where $x$ and $y$ are the numbers of photon detectors and single photon states. The single photon generation has always been an active research field, for both quantum communication and quantum metrology, and this has been widely studied using quantum dots[33,34], in order to reach high levels of purity, indistinguishability, and efficiency. For realistic photon detectors that do not have photon number resolving capability, such as on-off detectors, if the events of more than 1 photons arriving at the detectors are far less possible than those of only 1 photon, on-off detectors can be used as single photon detectors without introducing much problem. This is true for Method 4, since there are only two possibilities at the detector in each entanglement generator: either no photons detected, or a single photon detected. However, for Method 3, there is a relatively high probability that 2 photons arrive at one single detector since $BS_1$ and $BS_2$ are optimized chosen to reduce 2 photons off from the system. In this case, there is always a trade-off problem, where one can increase the BS transmissivity to reduce the probability of more than 1 photon appearing at detectors, with the cost of decreased generation efficiency of NOON states. Recently, photon-number-resolving detectors have been reported to be experimentally achievable with high efficiency[35]. With these developments, the experimental demonstration of the proposed theoretical work would be promising.

In conclusion, Method 4 has the highest efficiency under the condition that strong nonlinearity is achievable. Among the three linear methods, Methods 1 and 2 have higher generation probability compared with Method 3, where the latter is relatively more efficient. However, the Fock state input required for Method 2 makes it less feasible than Method 1 which uses coherent states.

**Data availability.**   No datasets were generated or analysed during the current study.

## Author Contributions

## Additional Information

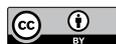 **Open Access** This article is licensed under a Creative Commons Attribution 4.0 International License, which permits use, sharing, adaptation, distribution and reproduction in any medium or format, as long as you give appropriate credit to the original author(s) and the source, provide a link to the Creative Commons license, and indicate if changes were made. The images or other third party material in this article are included in the article's Creative Commons license, unless indicated otherwise in a credit line to the material. If material is not included in the article's Creative Commons license and your intended use is not permitted by statutory regulation or exceeds the permitted use, you will need to obtain permission directly from the copyright holder. To view a copy of this license, visit http://creativecommons.org/licenses/by/4.0/.

© The Author(s) 2018